\documentclass[english,PRB,reprint,superscriptaddress]{revtex4-2}
\usepackage[LGR,T1]{fontenc}
\usepackage{textcomp}
\usepackage[latin9]{inputenc}
\usepackage{amsmath}
\usepackage{amssymb}
\usepackage{graphicx}

\makeatletter
\renewcommand{\fnum@figure}{Fig. \thefigure}

\makeatother

\usepackage{babel}

\begin{document}

\title{Strain-engineering spin-valley locking effect in altermagnetic monolayer with multipiezo properties}
\author{Yuqian Jiang}
\address{College of Sciences, Northeastern University, Shenyang 110819, China}
\author{Xinge Zhang}
\address{College of Sciences, Northeastern University, Shenyang 110819, China}
\author{Haoyue Bai}
\address{College of Sciences, Northeastern University, Shenyang 110819, China}
\author{Yuping Tian}
\address{College of Sciences, Northeastern University, Shenyang 110819, China}
\author{Binyuan Zhang}
\address{College of Sciences, Northeastern University, Shenyang 110819, China}
\author{Wei-jiang Gong}
\address{College of Sciences, Northeastern University, Shenyang 110819, China}
\author{Xiangru Kong}
\email{Contact author: kongxiangru@mail.neu.edu.cn}
\address{College of Sciences, Northeastern University, Shenyang 110819, China}

\begin{abstract}
Recently, altermagnetism (AM) in condensed matter systems has attracted much attention due to the physical properties arising from the alternating spins in both real space and reciprocal space.
In our work, we propose a stable monolayer Janus Nb$_2$SeTeO with altermagnetic ground state and a new type of spin-valley locking (SVL) effect. The monolayer Janus Nb$_2$SeTeO exhibits a mutipizeo effect with a large out-of-plane piezoelectricity and piezovalley effect with large valley polarization. 
The piezovalley effect is induced by the uniaxial strain effect in different directions, which contributes the anomalous valley Hall effect (AVHE) in the AM system. Moreover, the compressive uniaxial strain could induce the quantum anomalous Hall effect (QAHE) in the AM system, where the chirality of the dissipationless topological edge states could be manipulated by the direction of uniaxial strain. These manifest topological phase transitions could be realized via designing SVL effect in the AM system. Furthermore, the AM quantum spin Hall effect (QSHE) could be induced by the biaxial strain effect, which contributes the quantized spin Hall conductance. Our work reveals that by designing the SVL effect in mutipizeo Nb$_2$SeTeO monolayer could emerge new physics in AM systems, such as AVHE, QAHE and QSHE. 

\end{abstract}

\maketitle

\section{INTRODUCTION}
Altermagnetism (AM) is a new form of collinear magnetism that combines the physical properties of both antiferromagnets (AFM) and ferromagnets (FM)~\cite{Bai2024,mejkal2022,Cheong2024,PhysRevX.12.040501}. The spins in the AM system align antiparallel in real space as in the AFM system, but with nonrelativistic spin-splitting band structures as that in the FM system~\cite{PhysRevX.12.031042,PhysRevX.12.040002}. Therefore, the intrinsic properties of AM system overcome the disadvantages of weak magnetic response of AFM system and poor performance of FM spintronic devices~\cite{Bai2024}. Before the term AM was proposed~\cite{PhysRevX.12.040002}, the anomalous transport properties in AFM system were discovered including crystal Hall effect~\cite{mejkal2020}, momentum-dependent spin splitting~\cite{Hayami2019,PhysRevB.102.014422} and unconventional magneto-optics effect~\cite{Mazin2021}. Though the realistic spin-orbital coupling (SOC) effect is usually ignored in AM system, topological phases that induced by SOC effect could still exist in such systems~\cite{PhysRevB.109.024404,PhysRevB.109.L201109,parshukov2024topologicalresponsesgappedweyl,Li_2024,Wu2023}. Topological matter is becoming one of the popular fields in condensed matter physics due to the robust edge states, which could possibly be applied in future topological quantum computation. However, it is still rare to report the topological insulating phases in AM systems, which could facilitate the development of AM and spintronics.

In valleytronics, a new quantum degree of freedom is introduced as valley to investigate the electronic properties in the local extrema of energy band structures~\cite{Luo2024,Schaibley2016,Rycerz2007,PhysRevLett.99.236809,PhysRevLett.108.196802}.
Recently, spin-valley locking (SVL) effect has been proposed to investigate the dynamical physical processes that could benefit the realization of multifunctional materials~\cite{Liu2021,Ma2021,Zhu2019,PhysRevB.103.115401,Wang2020,Duprez2024,zhang2024crystalsymmetrypairedspinvalleylockinglayered}. The SVL effect induced by strong SOC effect and time-reversal symmetry is referred to as the $T$-paired SVL effect, where the sign in the spin splitting of the valleys is opposite. $T$-paired SVL effect exists in non-magnetic materials such as transition metal dichalcogenides. In AFM or AM systems, the SVL effect is referred to as the $C$-paired SVL effect as SOC effect is not required and the spin splitting of the valleys is controlled by the crystal symmetry. Therefore, the breaking of crystal symmetry could manipulate the valley splitting and valley polarizations~\cite{Wang2021}.
Compared with $T$-paired SVL effect, $C$-paired SVL effect could behave with long spin depolarization time and relate to multiple functions which promises its future applications~\cite{Ma2021}.

To study the new type of SVL effect in 2D materials, a monolayer of V$_2$Se$_2$O with antiferromagnetic order was proposed and investigated in Ref.~\cite{Ma2021}, and its bulk form is experimentally accessible~\cite{PhysRevB.98.075132}.
The multipiezo effect could also be achieved in the monolayer Janus V$_2$SeXO (X=Te for example)~\cite{Zhu2023}. In our work, we will investigate the strain engineering topological properties that associated with the SVL effect in this family of materials, and take monolayer Janus Nb$_2$SeTeO as an example. Our work demonstrates that the valley polarization and topological edge states could be engineered by the strain effect. The anomalous valley Hall effect (AVHE), quantum anomalous Hall effect (QAHE) and quantum spin Hall effect (QSHE) could all be realized in the AM monolayer Janus Nb$_2$SeTeO by controlling the directions of strain effect. 

\section{COMPUTATIONAL METHODS}

On the basis of density functional theory (DFT)~\cite{PhysRev.136.B864,PhysRev.140.A1133}, Vienna \textit{ab-initio} simulation package (VASP)~\cite{PhysRevB.54.11169,Kresse1996} by projector augmented wave (PAW) methods is utilized for the first-principles calculations. The generalized gradient approximation of Perdew-Burke-Ernzerhof (PBE GGA)~\cite{PhysRevLett.100.136406,PhysRevLett.77.3865} is chosen as the exchange correlation functional. The convergence criteria for energy and force are set at 10$^{-8}$ eV and 0.001 eV/\AA, respectively. The cut-off energy of the plane wave is set to
600 eV. To be consistent with the band gap calculated by HSE06 hybrid functional~\cite{Heyd2003}, The PBE+U
method~\cite{PhysRevB.52.R5467} is used to test different U values for the \textit{d} orbitals in Nb atoms. U is chosen as 4.6 eV in the methods proposed by Dudarev \textit{et al}~\cite{PhysRevB.57.1505}. 
A 24~\AA~vacuum layer is placed along the c-axis to avoid interlayer interactions. The Monkhorst-Pack (MP) grid is selected as $\Gamma$-centered
10 \texttimes{} 10 \texttimes{} 1~\cite{PhysRevB.13.5188}. 
The spectrum for the phonon dispersion was calculated to verify the dynamic stability using the Phonopy code~\cite{Togo2015} by the finite displacement method with a 3 \texttimes{} 3 \texttimes{} 1 supercell. N\'eel temperature is estimated by the Monte Carlo simulations implemented in the MCsolver package~\cite{LIU2019300}. 
The maximally localized Wannier function is obtained via the Wannier90 package~\cite{Mostofi2014}, and the topological properties are analyzed by the WannierTools package~\cite{Wu2018}. 

\section{RESULTS AND DISCUSSIONS}

\begin{figure}
\includegraphics[width=8.5cm,totalheight=6.375cm]{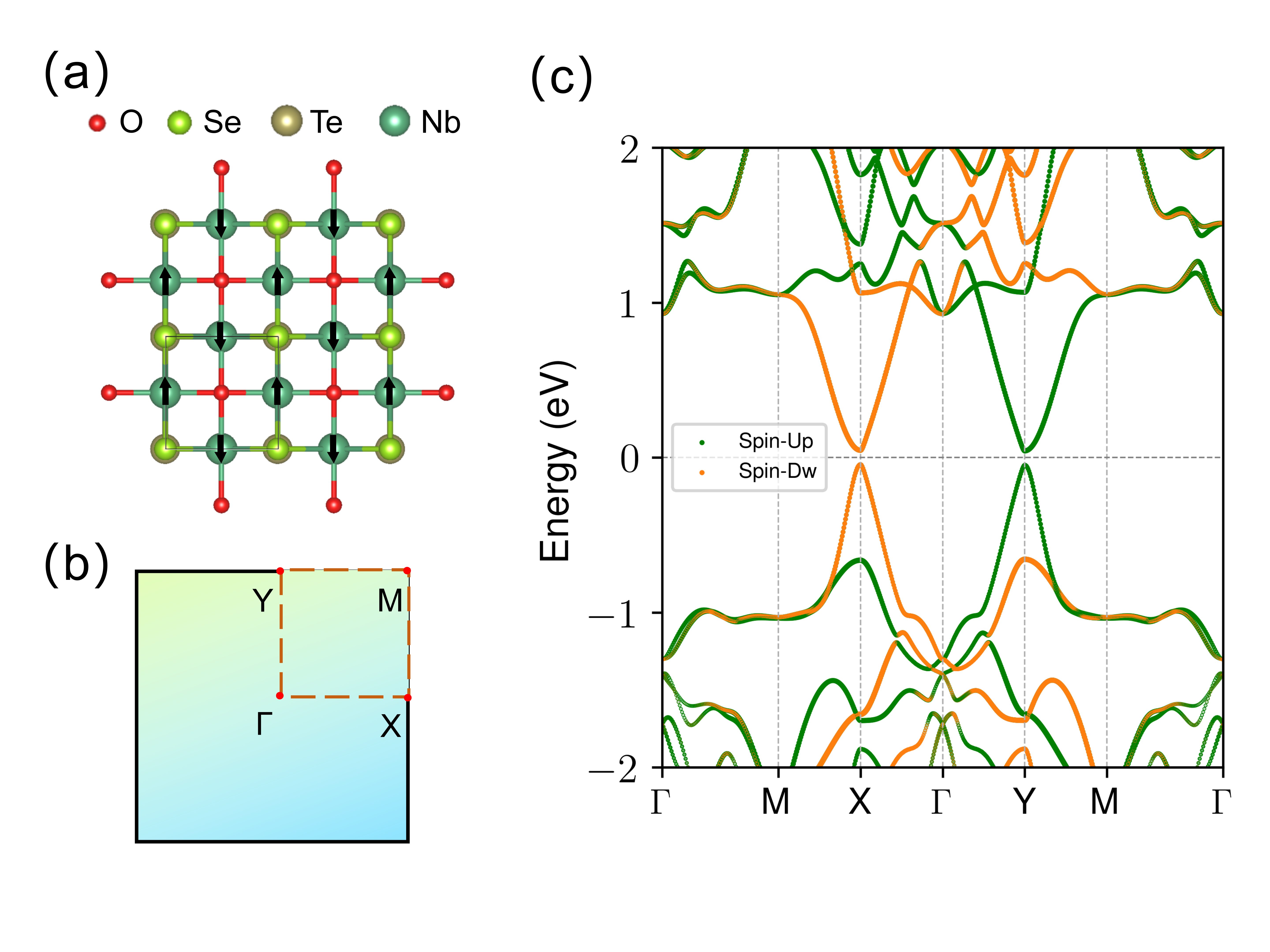}
\caption{(a) The top view of the monolayer Janus structure Nb$_2$SeTeO. Different spheres correspond to different atoms.
(b) The Brillouin zone with high symmetry points.
(c) The spin-polarized band structures with spin-up and spin-down (spin-dw) bands splitted in reciprocal space.}
\label{fig1}
\end{figure}

As shown in Fig.~\ref{fig1}, the monolayer Janus structure Nb$_2$SeTeO behaves as the two-dimensional (2D) square lattice pattern. 
The plane consistents of the Nb and O atoms are sandwiched between the top Se and the bottom Te planes as shown in Fig. S1(a) in the supplementary materials (SM), similar to V$_2$SeTeO~\cite{Zhu2023}. 
In order to confirm the magnetic ground state, four magnetic patterns of the 2 \texttimes{} 2 \texttimes{} 1 supercell of Nb$_2$SeTeO are constructed in Fig. S2.
The calculations for the energy of the four magnetic patterns indicate that the AM state in Fig.~\ref{fig1}(a) is the most stable. The energy for AM pattern is 2.03 eV lower than that of FM pattern.
By analyzing the magnetic exchange parameters of the AFM ground state, the N\'eel temperature can reach up to 200 K, as shown in Fig. S1(d).
By fully relaxing the structures with AM state via the first-principles calculations, the lattice constant is 4.12~\AA.
The dynamical stability of AM Nb$_2$SeTeO is confirmed by the phonon spectrum without imaginary frequency as shown in Fig. S1(b).
The elastic constants also satisfy Born-Huang criteria: C$_1$$_1$>0, C$_6$$_6$>0, |C$_1$$_1$|>|C$_1$$_2$|, where C$_1$$_1$=167.152 N/m, C$_1$$_2$=7.954 N/m, C$_6$$_6$=51.988 N/m. 
This also confirms the stability of monolayer Janus Nb$_2$SeTeO in AM state.

Due to the breaking horizontal mirror symmetry in monolayer Janus Nb$_2$SeTeO, there will also be a large out-of-plane piezoelectricity as in V$_2$SeTeO~\cite{Zhu2023}.
The piezoelectric tensor is defined as e$_{ijk}$ = $\partial$P${_i}$/$\partial$$\epsilon$$_{jk}$, in which P${_i}$ is the polarization and $\epsilon$$_{jk}$ is the strain effect.
The only one independent piezoelectric component e$_{31}$ will be 0.1196$\times$10$^{-10}$ c/m, with electrons contributing -0.1979$\times$10$^{-10}$~c/m and ions contributing 0.3175$\times$10$^{-10}$~c/m. 
The value is larger than that of WSTe, MoSTe and WSeTe~\cite{Dong2017}, making Nb$_2$SeTeO promising for future multifunctional applications.
The electronic band structure calculated by HSE06 in Fig.S1(c) demonstrate that the band gap is about 96 meV, which is consistent with that by PBE+U calculations without SOC as shown in Fig.~\ref{fig1}(c).
The band structure is spin polarized with bands splitted in reciprocal space.
This confirms the AM property in monolayer Janus structure Nb$_2$SeTeO with collinear AFM state~\cite{Hayami2019}.
As discussed in Ref.~\cite{Zhu2023}, the AM state could be elaborated from the broken symmetry rules.
In addition, there exists two valleys with opposite spin channels around $X$ and $Y$ valleys originating from the Nb atoms.
The SVL effect indicates a new type of valleytronics in AM square lattices, which promises possible applications for valleytronic materials.

\begin{figure}
\includegraphics[width=8.8cm,totalheight=6.6cm]{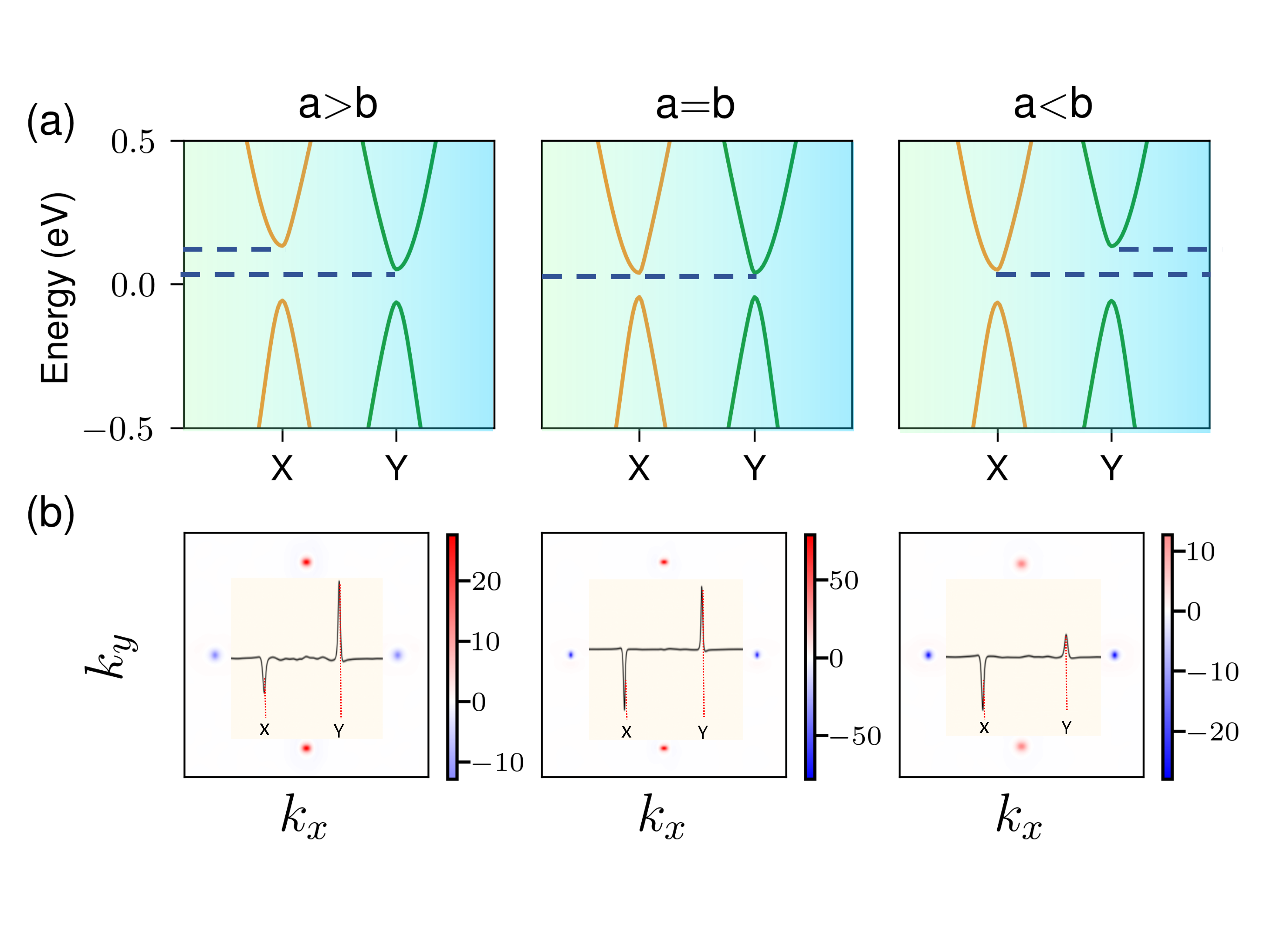}
\caption{(a) The spin-polarized band structures under uniaxial strain. $a$=$b$ indicates the unstrain condition, $a$>$b$ indicates the tensile strain is applied to the axis $a$, and $a$<$b$ indicates the tensile strain is applied to the axis $b$.
(b) The corresponding distributions of Berry curvature in reciprocal space.
The strength of the tensile strain is considered to be 2\%.}
\label{fig2}
\end{figure}

In previous studies~\cite{Zhu2023}, SOC effect is ignored as the SVL effect could be preserved. 
However, the strength of SOC effect should be considerable in Nb$_2$SeTeO due to the presence of heavy elements such as Se or Te. 
Under uniaxial strain effect, there will be valley polarization between the valleys around $X$ and $Y$ points, which can be termed as piezovalley effect such as in V$_2$SeTeO~\cite{Zhu2023}. 
In our work, we investigate the piezovalley effect with SOC effect in monolayer Janus Nb$_2$SeTeO by applying uniaxial strain effect. Both tensile (positive) and compressive (negative) strain are considered to investigate the uniaxial strain-engineering electronic structures. By applying the compressive uniaxial strain along $a$ or $b$ direction, monolayer Janus Nb$_2$SeTeO will preserve to be an insulating phase (Fig. S3(a)). However, there will be valley polarization between $X$ and $Y$ valleys as the valleys respond differently to the uniaxial strain in different directions.
Without any strain effect, the valleys at $X$ and $Y$ points are degenerate as shown in Fig.~\ref{fig2}(a).
Under the uniaxial tensile strain along $a$ direction or in the case of the lattice constant $a>b$, the energy of conduction bands at $X$ and $Y$ points will deviate and the energy at $X$ point will be higher than that at $Y$ point (Fig.~\ref{fig2}(a) and Fig. S3(d)).
However, in the case of $a<b$ or the uniaxial tensile strain along $b$ direction, the energy at $Y$ point will become higher than that of $X$ point.
The valley polarization induced by uniaxial strain indicates the piezovalley effect in monolayer Janus Nb$_2$SeTeO, which could be promising for the applications in multipizeo effect.
It could be observed that the valley polarization between the conduction bands increases monotonically versus the strength of uniaxial tensile strain effect, as shown in Fig. S4(a) and (c).

The valley polarization induced by the tensile strain effect signals the AVHE in monolayer Janus Nb$_2$SeTeO~\cite{Tong2016}.
The AVHE triggers a new area of valleytronics that requires breaking time-reversal symmetry and SOC effect.
Usually, AVHE is proposed in ferromagnetic monolayer systems such as FeCl$_2$~\cite{Hu2020}, RuCl$_2$~\cite{PhysRevB.105.195312}, and TiTeI~\cite{PhysRevB.107.174434}.
However, AVHE could also be realized in AFM monolayers such as MnPSe$_3$ interfaced with a ferroelectric Sc$_2$CO$_2$ monolayer~\cite{Du2022}.
To elucidate the AVHE in AM monolayer Janus Nb$_2$SeTeO, Berry curvature is derived from the equation:
\begin{equation}
\begin{aligned}
\Omega(\boldsymbol{k}) & =-\sum_{n}\sum_{n\ne m}f_{n}(\boldsymbol{k})\\
 & \times 2\operatorname{Im} \frac{<\psi_{n\boldsymbol{k}}\mid\hat{v}_{x}\mid\psi_{m\boldsymbol{k}}><\psi_{m\boldsymbol{k}}\mid\hat{v}_{y}\mid\psi_{n\boldsymbol{k}}>}{(E_{n\boldsymbol{k}}-E_{m\boldsymbol{k}})^{2}},
\end{aligned}
\end{equation}
where $\boldsymbol{k}$ being the electron wave
vector and $f_{n}(\boldsymbol{k})$ defines as the Fermi-Dirac
distribution function, $\hat{v}_{x}$ and $\hat{v}_{y}$ are
velocity operators of the Dirac electrons, \textit{n} and \textit{m} are the band indexes, $E_{n\boldsymbol{k}}$ and $E_{m\boldsymbol{k}}$ are the eigenvalues of the Bloch wave functions $\psi_{n\boldsymbol{k}}$ and $\psi_{m\boldsymbol{k}}$, respectively.
As shown in the middle panel of Fig.~\ref{fig2}(b), in the absence of any strain effect ($a=b$), the values of Berry curvature peaks around $X$ and $Y$ valleys are almost equal, but with opposite signs.
In the case of $a$>$b$ ($a$<$b$) with tensile strain along $a$ ($b$) direction, the value of Berry curvature peaks around $Y$ ($X$) valleys will be larger than that around $X$ ($Y$) valleys, as shown in the left and right panels of Fig.~\ref{fig2}(b), respectively.
The unbalanced Berry curvature between $X$ and $Y$ valleys indicates a nonzero anomalous Hall conductivity (AHC), which could be calculated from below:
\begin{equation}
\sigma_{xy}=-\frac{e^{2}}{\hbar}\int_{BZ}\frac{d^{3}k}{(2\pi)^{3}}\Omega(\boldsymbol{k})
\end{equation}
where BZ defines as the first Brillouin zone, $\boldsymbol{k}$ being the electron wave vector, $\Omega(\boldsymbol{k})$ is Berry curvature.
The result is presented in Fig.~S5.
This confirms the existence of tensile strain-induced AVHE in monolayer Janus Nb$_2$SeTeO, which signals the promising applications in valleytronics.

\begin{figure*}
\includegraphics[width=1.85\columnwidth]{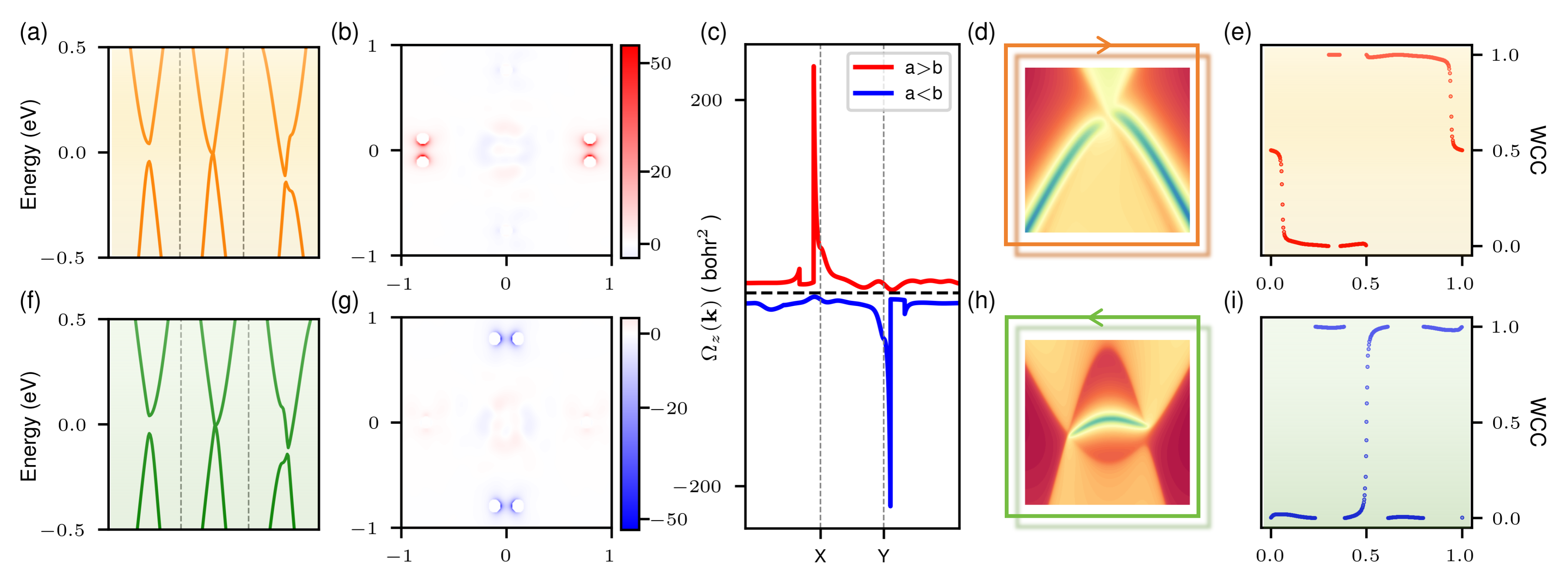}
\caption{The topological phase transitions induced by compressive uniaxial strain. The band inversions around (a) $X$ valley for the uniaxial strain along $a$ and (f) $Y$ valley for for the uniaxial strain along $b$. (b,g) The corresponding distributions of Berry curvature in reciprocal space.
(c) The corresponding distributions of Berry curvature around $X$ and $Y$ valleys.
(d,h) The corresponding strain-induced chiral edge states.
(e,i) The evolution of Wannier Charge Centers (WCCs).
}
\label{fig3}
\end{figure*}

On the other hand, if the uniaxial strain is compressive, topological phase transitions could be induced and switchable chiral edge states could be obtained. The main results are presented in Fig.~\ref{fig3}. The compressive uniaxial strain induced band inversions occur around $X$ or $Y$ valleys in the strength of -1.2\%. As demonstrated in Fig.~\ref{fig3}(a) and (f), the compressive uniaxial strain along $a$ will induce the band closing and reopening around $X$ valley, while the compressive uniaxial strain along $b$ will induce the similar phenomena around $Y$ valley, and the result is also shown in Fig. S4(b) and (e). The strain-induced band gap will shift downwards in relation to the Fermi level. But the system itself will preserve to exhibit local gap in the whole Brillouin zone, which is consistent with the band structures in Fig. S3 (c) and (f) in the strength of -4\%. By increasing the strength of compressive uniaxial strain to about -5.8\%, the electronic states in Nb$_2$SeTeO will become metallic. The spin-resolved bands in the metallic state are still split as shown in Fig. S3 (b) and (e). This indicates that the compressive uniaxial strain does not affect the AM properties. Therefore as shown in Fig. S3 (a), the AM metallic phase induced by the compressive strain larger than -5.8\% is termed as AM-metal, and the AM insulating phase induced by tensile strain and compressive strain smaller than -1.2\% is termed as AM-insulator, as shown in Fig. S3 (d) and (g). In the range of -5.8\% to -1.2\% for the strength of compressive strain, QAHE could be induced by the strain effect because the band inversion could only appear in one spin channel. As shown in Fig.~\ref{fig3}(a) and (f), the band inversion in the spin-up channel is induced by uniaxial strain along $a$, while the band inversion in the spin-down channel is induced by uniaxial strain along $b$. The strain-induced QAHE is termed as AM-QAHE as shown in Fig. S3(a).

To demonstrate the topological properties in AM-QAHE, Berry curvatures under uniaxial strain in different directions are analyzed in Fig.~\ref{fig3} (b), (c) and (g).
Under uniaxial strain along $a$ direction (or $a>b$), the Berry curvatures for the occupied states around $Y$ point are absent, while those around $X$ point are large and positive values (Fig.~\ref{fig3} (b) and (c)). As discussed above, the band inversions induced by the compressive strain effect actually originate from piezovalley properties. Thus, under uniaxial strain along $b$ direction (or $a<b$), the Berry curvatures for the occupied states will behave oppositely to those in the case of $a>b$. The Berry curvatures for the occupied states around $X$ point will be absent, while those around $Y$ point will be large and negative values (Fig.~\ref{fig3} (b) and (g)). The unbalanced Berry curvatures between $X$ and $Y$ valleys will usually result in nonzero AHC in Eq. 2. However, in the framework of topological phase transitions, the AHC could be quantized, giving the nonzero integers, i.e., Chern number $C$. The Chern number could be confirmed by the evolution of WCCs as shown in Fig.~\ref{fig3} (e) and (i).
The downwards shift of WCCs in Fig.~\ref{fig3} (e) indicates the Chern number $C=-1$, which corresponds to a left-moving chiral edge state in Fig.~\ref{fig3} (d).
Similarly, the upward shift of WCCs in Fig.~\ref{fig3} (h) indicates the Chern number $C=1$, which corresponds to a right-moving chiral edge state in Fig.~\ref{fig3} (h).
The chiral edge states in QAHE are robust to both nonmagnetic and magnetic disorders, contributes to the dissipationless conductance channels.
So far, though QAHE has been reported in Cr-doped Bi$_2$Te$_3$ thin films~\cite{Chang2013} and MnBi$_2$Te$_4$~\cite{Deng2020}, there are still difficulties in its experimental realization.
Either the experimental temperature is low or an external magnetic field is required.
Therefore, it is necessary to engineering QAHE on purpose via manipulations.
The results of our work demonstrate that the dissipationless chiral edge states in AM monolayer Janus Nb$_2$TeSeO could be manipulated by compressive strain effect.
The reversal of the flow direction for the chiral edge states could be tuned by the uniaxial strain effect in two orthogonal directions.
This could provide as input or output signals for future quantum devices.

\begin{figure*}
\includegraphics[width=2.0\columnwidth]{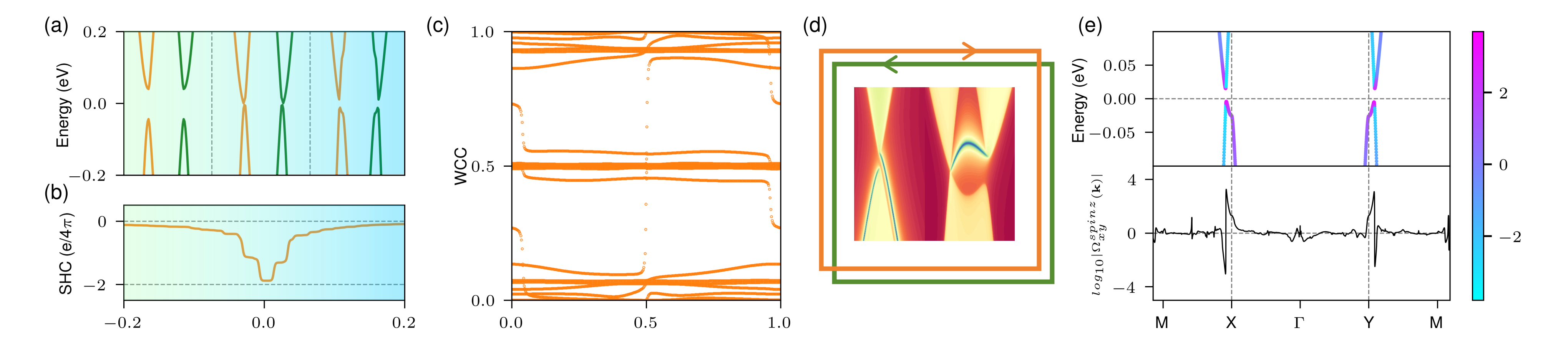}
\caption{The topological phase transitions and topological properties induced by biaxial strain.
(a) The band inversions appear simultaneously around both $X$ and $Y$ valleys.
The quantized spin Hall conductance (SHC) in the topological phase (b), the evolution of WCCs (c) and the corresponding chiral edge states with opposite chirality (d).
(e) The band-resolved (top panel) and the $k$-resolved (bottom panel) spin Berry Curvature like term.
}
\label{fig4}
\end{figure*}

Furthermore, biaxial strain effect is also considered to engineer the physical properties in monolayer Janus Nb$_2$SeTeO.
Under the tensile biaxial strain, the band gaps around both $X$ and $Y$ valleys will increase linearly versus the strength of biaxial strain effect, as shown in Fig. S7.
Moreover, the band gap between $X$ and $Y$ valleys will behave the similar trends, i.e., they will nearly be equal under any strength of biaxial strain effect.
As shown in Fig. S6(e), under the strength of biaxial strain with 10\%, the system will exhibit large gap insulating properties.
So the phase is termed as AM-insulator but without valley polarization, compared to the case under tensile uniaxial strain effect as shown in Fig. S3(a).
However, under the compressive biaxial strain effect, the band gaps around $X$ and $Y$ valleys will firstly decrease and then band inversions appear.
The process of band inversions versus the biaxial strain is demonstrated in Fig.~\ref{fig4} (a) and Fig. S7.
It could be observed that the band gaps around $X$ and $Y$ valleys will close and reopen simultaneously in the strength of 1.2\% for the compressive biaxial strain, which indicates the band inversions appear simultaneously in both spin-up and spin-down channels.
It is known that uniaxial strain leads to a topological phase transition in only one spin channel with chiral edge states in opposite Chern numbers.
Therefore, around $X$ and $Y$ valleys, there will exist two chiral edge states with opposite flowing directions as shown in Fig.~\ref{fig4} (d).
The two chiral edge states will not contribute the AHC, but will result in quantized spin Hall conductivity (SHC) as shown in Fig.~\ref{fig4} (b).
The SHC is calculated by the following formula:
\begin{equation}
\begin{split}
\sigma_{xy}^{s_z}(\omega) & =\hbar\int_{BZ}\frac{d^{3}k}{(2\pi)^{3}}\sum\limits_{n}f_{n\boldsymbol{k}}\\
 & \times\sum\limits_{m\ne n}\frac{2\operatorname{Im}[<n\boldsymbol{k}|\hat{j}_{x}^{s_z}|m\boldsymbol{k}><m\boldsymbol{k}|-e\hat{v}_{y}|n\boldsymbol{k}>]}{(\epsilon_{n\boldsymbol{k}}-\epsilon_{m\boldsymbol{k}})^{2}-(\hbar\omega+i\eta)^{2}},    
\end{split}
\end{equation}
where \textit{n} and \textit{m} are the band indexes, $\epsilon_{n}$ and $\epsilon_{m}$ are the eigenvalues, $\boldsymbol{k}$ being the electron wave vector, $f_{n\boldsymbol{k}}$ is the Fermi distribution function, $\hat{j}_{x}^{s_z}$ is the spin current operator in the projection of spin $z$ direction ($s_z$), $\hat{v}_{y}$ is the velocity operator and both the angular frequency $\omega$ and the damping factor $\eta$ are reduced to zero in the scenario of a direct current with a clean limit.
The nearly quantized SHC indicates QSHE could exist in time-reversal symmetry breaking AM system, though the two chiral edge states in Fig.~\ref{fig4}(d) do not form helical edge states as in the time-reversal symmetry protected QSHE.
In Fig.~\ref{fig4}(e), the band-resolved and $k$-resolved spin Berry curvature like term concludes the SHC in Fig.~\ref{fig4}(b) is mainly contributed by the spin Berry curvature like term around $X$ and $Y$ valleys.
In our recent work~\cite{Tian2024}, a new type of layer-coupled QSHE is reported as quantum layer spin Hall effect, which is also characterized by two separate chiral edge states and quantized SHC.
In this work, we refer to the topological phase as AM-QSHE, as shown in Fig. S6(d).
The continuity in the evolution of WCCs for the whole Brillouin zone in Fig.~\ref{fig4}(c) also reveals the topological nontrivial properties in the biaxial strain-induced AM-QSHE, which is similar to that in Ref.~\cite{Tian2024}.

As the strength of compressive strain effect increases, the band gaps in the AM-QSHE will disappear and a metallic state will be formed as shown in Fig. S6(a).
The bands in different spin channels is still splitted as shown in Fig.~S6(c), which characterizes the AM properties.
So it could be termed as AM-metal phase.
After the strength of compressive strain effect exceeds 10.2\%, the bands in different spin channels will not be splitted as shown in Fig. S6(b), so the monolayer Janus Nb$_2$SeTeO will not present AM properties and we just term this phase as the normal (NM) metal phase. 

Finally, the biaxial strain-induced AM-QSHE indicates a new type of topological phase in AM systems, which present quantized SHC.
It is also notable that the uniaxial strain will induce AM-QAHE but the chirality of topological edge states could be manipulated by the direction of uniaxial strain.
The AM-QSHE could be considered as two copies of AM-QAHE but with opposite Chern numbers.
Both AM-QSHE and AM-QAHE are related to the band inversions around $X$ and $Y$ valleys, thus correlating with the piezovalley and AM properties in monolayer Janus Nb$_2$SeTeO.

\section*{CONCLUSIONS}
In conclusion, through first-principles calculations, we predict that the monolayer Janus Nb$_2$SeTeO is an stable altermagnet and behaves multipiezo properties.
Especially, under uniaxial strain along $a$ or $b$ direction, the $X$ and $Y$ valleys could respond differently in the valley polarization, i.e., the piezovalley properties.
Thus, the AVHE could emerge in the AM system via the tensile uniaxial strain effect.
The compressive uniaxial strain causes band inversions around $X$ or $Y$ valleys, resulting in AM-QAHE.
The AM-QAHE is confirmed by the calculations of the distributions of Berry curvature in reciprocal space, chiral edge states and the evolution of WCCs.
The uniaxial strain along $a$ or $b$ direction will manipulate the flowing direction of the chiral edge states, thus could promise its applications in future quantum devices.
The tensile biaxial strain will enlarge the band gaps around both $X$ and $Y$ valleys, which results in the AM-insulator phase.
However, the compressive biaxial strain will simultaneously cause the band inversions around the $X$ and $Y$ valleys.
The AM-insulator will transition into AM-QSHE with quantized SHC and two chiral edge states with opposite Chern numbers around $X$ and $Y$ valleys.
The topological properties in AM-QSHE are similar as the quantum layer spin Hall effect reported previously.
Our work demonstrates that strain effect can be used to tune the physical properties in monolayer Janus Nb$_2$SeTeO on purpose, thus advancing the development and applications in spintronics and nanodevices.



\section*{ACKNOWLEDGMENTS}
X. K. acknowledges the start up funding from Northeastern University, China. 
B. Z. acknowledges the the National Science Foundation of China (Grant No. 12204091).
This work was financially supported by the LiaoNing Revitalization Talents Program (Grant No. XLYC1907033), the Natural Science Foundation of Liaoning province (Grant No. 2023-MS-072 and No. 2024-MSBA-36), and the Fundamental Research Funds for the Central Universities (Grant Nos. N2209005 and N2205015).

%

\end{document}